\documentclass[%
aps,
prx,
reprint,
superscriptaddress,
nofootinbib,
amsmath,
amssymb
]{revtex4-2}
\usepackage{
    physics,
    graphicx,
    multirow,
    tabularx,
    mathtools, %for \begin{rcases} \end{rcases}
    placeins, %To include \FloatBarrier
    nicefrac, %For inline fraction
    hyperref, %To use put in url, enable links in bibiliography
    threeparttable, % For table footnotes
    siunitx,
    enumitem,
    comment,
    physics,
    stmaryrd,
    amsthm
}

\usepackage[capitalize]{cleveref}
\Crefname{section}{Sec.}{Secs.}

\usepackage[dvipsnames]{xcolor}
\hypersetup{
    colorlinks,
    linkcolor={blue!50!black},
    citecolor={blue!50!black},
    urlcolor={blue!80!black}
}% remove the box around citation link and bibilio link

\usepackage[caption=false]{subfig} %The caption part is not compatible with revtex
\graphicspath{{Figures/}}

\NewDocumentCommand{\evalat}{sO{\big}mm}{%
  \IfBooleanTF{#1}
   {\mleft. #3 \mright|_{#4}}
   {#3#2|_{#4}}%
}

\newtheorem{theorem}{Theorem}

\newtheorem{remark}{Remark}[theorem]
\newtheorem*{proof*}{Proof}

\begin{document}

%\preprint{APS/123-QED}

\title{Algorithmic error mitigation for quantum eigenvalues estimation}

\newcommand{\oxddress}{\affiliation{Department of Materials, University of Oxford, Parks Road, Oxford OX1 3PH, United Kingdom}}
\newcommand{\qmaddress}{\affiliation{Quantum Motion, 9 Sterling Way, London N7 9HJ, United Kingdom}}
\newcommand{\osakauniaddress}{\affiliation{Graduate School of Engineering Science, Osaka University, Osaka 560-8531, Japan}}
\newcommand{\qiqbaddress}{\affiliation{Center for Quantum Information and Quantum Biology, Osaka University, 1-2 Machikaneyama, Toyonaka, Osaka 560-0043, Japan}}
\newcommand{\rikenaddress}{\affiliation{Center for Quantum Computing, RIKEN, Hirosawa 2-1, Wako Saitama 351-0198, Japan}}
\author{Adam Siegel}
\email{adam.siegel@materials.ox.ac.uk}
\oxddress
\qmaddress

\author{Kosuke Mitarai}
\osakauniaddress \qiqbaddress 

\author{Keisuke Fujii}
\osakauniaddress \qiqbaddress \rikenaddress

\date{\today}% It is always \today, today,
%  but any date may be explicitly specified

\begin{abstract}
    When estimating the eigenvalues of a given observable, even fault-tolerant quantum computers will be subject to errors, namely \textit{algorithmic errors}. These stem from approximations in the algorithms implementing the unitary passed to phase estimation to extract the eigenvalues, \textit{e.g.} Trotterisation or qubitisation. These errors can be tamed by increasing the circuit complexity, which may be unfeasible in early-stage fault-tolerant devices. Rather, we propose in this work an error mitigation strategy that enables a reduction of the algorithmic errors up to any order, at the cost of evaluating the eigenvalues of a set of observables implementable with limited resources. The number of required observables is estimated and is shown to only grow polynomially with the number of terms in the Hamiltonian, and in some cases, linearly with the desired order of error mitigation. Our results show error reduction of several orders of magnitude in physically relevant cases, thus promising accurate eigenvalue estimation even in early fault-tolerant devices with limited number of qubits.
\end{abstract}

\maketitle

\section{Introduction}

The eigenvalues of observables characterising a quantum system serve as a critical benchmark for understanding the system's behaviour and properties. Accurate estimation of the eigenvalues, such as the ground-state energy of the Hamiltonian, enables researchers to unravel key phenomena like chemical reactions, material properties, and phase transitions. While the task of estimating the eigenvalues becomes exceedingly challenging for classical computers as the size of the quantum system increases, quantum computers offer a promising avenue for tackling this problem efficiently, in particular with the use of the well-known phase estimation algorithm (PE) \cite{kitaev1995quantum, O_Loan_2009, Kimmel_2015, Wan_2022}. No matter what variant of the algorithm is used, PE always takes as input a unitary operator $\mathcal{W}(A)$ whose eigenvalues are a function of the eigenvalues of the target observable $A$. In the case of the Hamiltonian, most past work have focused on Hamiltonian Simulation, i.e. $\mathcal{W}(H)=\mathrm{e}^{-\mathrm{i}Ht}$ for a given time $t$, which can be implemented by Trotterisation \cite{Hatano_2005, yi2021spectral} or truncated Taylor series \cite{Berry_2015}. More recent papers \cite{Berry_2018, Poulin_2018} have however advocated for the implementation of $\mathcal{W}(H)=\mathrm{e}^{-\mathrm{i}\mathrm{arccos}(H/\lambda)}$ with $\lambda$ a parameter greater than the spectral norm of $H$. This unitary can besides be implemented efficiently by qubitisation \cite{Low_2019}.

Nonetheless, all these methods are subject to errors \textit{even in fault-tolerant quantum computers} where gate errors can be neglected. These leftover errors, called \textit{compilation errors} or \textit{algorithmic errors}, stem from approximations the algorithm uses to implement the unitary operator $\mathcal{W}(A)$: in the case of Trotterisation, the exponential of a sum is broken down into a product of exponentials, which is only exactly true when all terms of the sum commute; as to qubitisation, the protocol requires the preparation of a state $\ket{G}$ encoding information about the target Hamiltonian, which can only be executed up to a certain error \cite{Babbush_2018}. These algorithmic errors can of course be tamed by augmenting the circuit complexity within the quantum algorithm, thereby improving its precision, but this may be unfeasible in early-stage fault-tolerant quantum computers where the number of qubits and gates will have to remain moderate (in particular non-Clifford gates \cite{akahoshi2023partially}). Another solution is the use of error mitigation techniques, which consist of attenuating the noise by the means of classical post-processing. These techniques include zero-noise extrapolation \cite{Endo_2019, Endo_2021, Otten_2018}, purity constraints \cite{Huggins_2021} or subspace expansion \cite{McClean_2017}.

In this paper, we present an error mitigation technique targeting algorithmic errors, which can be applied not only for the usual estimation of expectation values, but also for eigenvalues. To this end, our method expands on Richardson extrapolation \cite{Temme_2017}, with key novel elements: if this extrapolation has already been identified as a promising candidate for error mitigation in the context of Hamiltonian Simulation \cite{Endo_2019, Vazquez_2022}, or used in conjunction with error correction \cite{gonzales2023fault, wahl2023zero, Suzuki_2022}, our contribution lies in generalising it. More precisely, we first extend it to the estimation of eigenvalues, which requires the use of precise and non-trivial mathematical theorems. Second, we enable its application to the most optimal Hamiltonian Simulation algorithm, namely qubitisation, by employing a multi-parameter formalism introduced later in the paper.
% introduce a multi-parameter formalism and meticulously study its optimisation, to render our applicable to the most optimal Hamiltonian Simulation algorithm, namely qubitisation.
% , including its extension to eigenvalues estimation. Even if already understood to provide an advantage in the context
% We first focus on the estimation of eigenvalues rather than expectation values, which is a natural but mathematically non-trivial extension of the aforementioned error mitigation schemes. In order to make it applicable to a wide 
% We besides focus on algo errors ie fully known
% Unlike the aforementioned error mitigation schemes, ours focuses on  eigenvalues estimation and relies on the full knowledge of the errors affecting the system.
By implementing multiple imperfect but fully determined unitaries and extracting their eigenvalues, we show that we can build an estimate of the target eigenvalue up to any order $p$. To assess the effectiveness of this method, we conduct numerical tests and observe a significant reduction in errors, reaching several orders of magnitude, such as in the case of Trotter errors. The overall cost of the procedure is also evaluated and is shown to remain particularly modest as one only has to independently perform phase estimation a sufficient number of times. To correct up to $p$-th order, we estimate this number to grow as a polynomial of degree $p$ with the number of terms of the Hamiltonian, or even linearly with $p$ in some relevant cases such as Trotterisation.
% In this paper, we present a novel error mitigation technique for the reduction of algorithmic errors. We first describe the mitigation protocol and show that it can theoretically suppress errors up to any order. Then, we present numerical simulations that confirm the efficiency of the proposed method and estimate its cost.

\section{Methods}

In this section, we present our error mitigation scheme. After stating the assumptions it relies on, we describe the error mitigation protocol and theoretically estimate its performance.

\subsection{Algorithmic errors of observables}

We are interested in estimating the eigenvalues of a target observable $A$ on a fault-tolerant computer. This can be done with the use of the phase estimation algorithm (PE), assuming that a unitary operator $\mathcal{W}(A)$ can be implemented. However, owing to limited circuit depth for the generation of such unitary operator, the implementation may only be performed up to some algorithmic (or compilation) errors. The consequence is the implementation of an actual unitary $\mathcal{W}(A')$, associated with an effectively implemented observable $A'$, whose distance to the target observable $A$ depends on the strength of the algorithmic errors.

One particularly important thing to note is that, because we are focusing on a fault-tolerant setup, $A$ and $A'$ are close to deterministic. Indeed, we can consider in a first approximation that the logical error rate is null in a FTQC and that only algorithmic errors exist. Consequently, one exactly knows the target observable $A$ as well as the circuit that generated the approximate observable $A'$ (chosen by the user). $A$ and $A'$ can then be parameterised by a functional $\tilde{A}(\delta_1,...,\delta_N)$ that satisfies:
\begin{align}
    A &=\tilde{A}(0,...,0)\\
    A'&=\tilde{A}(\delta_1,...,\delta_N)
\end{align}
for a specific choice of $\delta_i$'s which the represent algorithmic errors. Importantly, these $\delta_i$'s are entirely known.

This can be illustrated with two examples, in the case of the eigenvalue estimation of a Hamiltonian. We will thus note the target and implementable Hamiltonians $H$ and $H'$ respectively.
When $H=\sum H_k$ and $\mathcal{W}(H)=\mathrm{e}^{-\mathrm{i}Ht}$ is implemented by Trotterisation, the algorithmic errors stem from the inexact splitting of the exponential of a sum into a product of exponentials. The effective Hamiltonian under first-order Trotterisation is:
\begin{equation} \label{trotter}
    H'=\frac{\mathrm{i}}{\delta t}\mathrm{log}\left(\prod\mathrm{e}^{-\mathrm{i}H_k\delta t}\right)
\end{equation}
with $\delta t$ a small fraction of the total time $t$. As a result, the functional describing $H$ and $H'$ can be chosen to depend on a single parameter: $N=1$ and $\delta_1=\delta t$.

Likewise, if $H$ is expressed as a linear combination of unitaries $H=\sum c_iP_i$, and $\mathcal{W}(H)=\mathrm{e}^{-\mathrm{i}\mathrm{arccos}(H/\lambda)}$ is implemented by qubitisation \cite{Berry_2018, Babbush_2018}, then the protocol relies on the preparation of the state $\ket{G}=\sum\sqrt{c_i/\lambda}\ket{i}$ with $\lambda=\sum c_i$. More precisely, if the state $\ket{G}$ can be prepared, qubitisation enables an exact implementation of $\mathcal{W}(H)$ (assuming no gate errors). However, the state preparation itself is subject to algorithmic errors \cite{Babbush_2018} and a state $\ket{G'}=\sum\sqrt{c'_i/\lambda'}\ket{i}$ with $\lambda'=\sum c'_i$, may instead be generated. As a result, the effectively implemented observable is:
\begin{equation} \label{qubitisation}
    H'=\sum_i c'_iP_i
\end{equation}
The functional describing $H$ and $H'$ hence depends on multiple parameters $\delta_i=c'_i-c_i$ which, again, are known. For instance, in \cite{Babbush_2018}, $c'_i$ is the $\mu$-bit binary approximation of $c_i$, with $\mu$ related to the number of ancilla qubits used in the state preparation circuit.

\subsection{Error mitigation protocol}

The problem we consider here is the following: given a target observable $A$ and an implementable observable $A'$ that are both entirely known, how can we accurately estimate the eigenvalues of the target observable $A$ from the eigenvalues of the implementable observable $A'$? We show in this section that this task can be fulfilled efficiently provided not one, but multiple and distinct, observables $A'_k$ approximating $A$ are implemented:
$$A'_k=\tilde{A}(\delta_{1,k},...,\delta_{N,k})$$
From now on, we will denote by $m$ the number of implemented $A'_k$'s. $N$ will still designate the number of coefficients an individual $A'_k$ depends on.
In the above equation, the second subscript $k \in \llbracket 1,m\rrbracket$ of $\delta_{i,k}$ will refer to the implemented observable $A'_k$ while the first one $i \in \llbracket 1,N\rrbracket$ will refer to the specific coefficient. Alternatively, $(\delta_{1,k},...,\delta_{N,k})$ may be noted in a more compact form as an $N$-dimensional vector $\vec{\delta}_k$.

There are various ways of generating these $A'_k$ depending on the situation. Considering the two previous examples again: for Trotterisation, it is enough to vary the infinitesimal time step $\delta t$; for qubitisation, in particular within the state preparation protocol of \cite{Babbush_2018}, one can generate the closest $\mu$-bit binary numbers $c'_{i,k}$ to each of the target coefficients $c_i$ (rather than just one $c'_i$ per $c_i$).

In this setup, our main result reads as follows:

\begin{theorem}
    Let $A$ be an observable whose eigenvalues we want to estimate up to a given order $p$, and $(A'_k)$ be a set of observables, in the neighbourhood of $A$, whose eigenvalues we can estimate. Assume that the observables in the vicinity of $A$ can be described by a smooth functional $\tilde{A}(\delta_1,...,\delta_N)$ such that:
    \begin{align}
        A &= \tilde{A}(0,...,0),\\
        \forall k~~A'_k &= \tilde{A}(\delta_{1,k},...,\delta_{N,k}),
    \end{align}
    Assume that, for all $i$ and $k$, the coefficients $\delta_{i,k}$ parameterising $A'_k$ are known. It follows that one can construct coefficients $(\lambda_1,...,\lambda_m)\in\mathbb{R}^m$ such that any non-degenerate eigenvalue $a$ of $A$ can be estimated up to order $p$ from eigenvalues $a'_k$ of $A'_k$ and from the coefficients $\lambda_k$:
    \begin{equation} \label{master}
        \exists m\in\mathbb{N}~~\exists (\lambda_1,...\lambda_m)\in\mathbb{R}^m~~
        a = \sum_{k=1}^m \lambda_k a'_k + \delta a
    \end{equation}
    where
    \begin{equation}
    \delta a = O\left(||\lambda||_1 ||\delta||_\infty^{p+1}\right)
    \end{equation}
    where $||.||_1$ denotes the $\ell^1$-norm and $||\delta||_\infty = \max_{i,k} |\delta_{i,k}|$.
    %     \textcolor{blue}{(Adam) ... parameterising $A'_k$ are known. It follows that one can construct coefficients $(\lambda_1,...,\lambda_m)\in\mathbb{R}^m$ such that any non-degenerate eigenvalue $a$ of $A$ can be estimated up to order $p$ from eigenvalues $a'_k$ of $A'_k$ and from the coefficients $\lambda_k$:
    % \begin{equation} \label{master}
    %     \exists m\in\mathbb{N}~~\exists (\lambda_1,...\lambda_m)\in\mathbb{R}^m~~
    %     a = \sum_{k=1}^m \lambda_k a'_k + \delta a
    % \end{equation}
    % where
    % \begin{equation}
    % \delta a = O\left(||\lambda||_1 ||\delta||_\infty^{p+1}\right)
    % \end{equation}
    % where $||.||_1$ denotes the $\ell^1$-norm and $||\delta||_\infty = \max_{i,k} |\delta_{i,k}|$. 
    The required number of coefficients (or equivalently of observables $A'_k$) satisfies:
    \begin{equation}
        m = O(N^p)
    \end{equation}
    % }

\end{theorem}

\begin{remark}
    If the functional $\tilde{A}$ depends on one parameter only ($N=1$), the eigenvalue $a$ we are trying to estimate need not be non-degenerate.
\end{remark}

\begin{proof}
    Let $a$ be a non-degenerate eigenvalue of $A$ and $p\in\mathbb{N}$ be the order up to which we want to estimate $a$. Since $a$ is non-degenerate, it is a locally totally differentiable function of the operator $A$ \cite{kato1995perturbation}. Alternatively, as per Remark 1.1, if $\tilde{A}$ depends on one parameter only, the eigenvalue $a$ need not be non-degenerate to be a totally differentiable function of this parameter \cite{kato1995perturbation}. In both situations and in less technical terms, this means that $a$ varies smoothly when $A$ is perturbed. Thus, if $A'$ is an observable in the neighbourhood of $A$ such that:
    \begin{equation*}
        A'=\tilde{A}(\delta_1,...,\delta_N)
    \end{equation*}
    then there exists an eigenvalue $a'$ of $A'$ in the neighbourhood of $a$, which can be expressed via a generalised Taylor expansion (using the notation $|i|=i_1+...+i_N$):
    % \begin{equation*}
    %     a' = a + \sum_{q=1}^p a'^{(q)} + O(||\delta||^{p+1})
    % \end{equation*}
    % where $a'^{(q)}$ is a $q$-th order correction:
    % \begin{align}
    %     a'^{(q)} &= \sum_{\scriptscriptstyle i_1+...+i_N=q}\delta_1^{i_1}...\delta_N^{i_N}\gamma_{i_1,...,i_N}\\
    %     \gamma_{i_1,...,i_N} &= \frac{1}{q!}{N\choose k}\evalat[\Big]{\frac{\partial^q a}{\partial\delta^{i_1}...\partial\delta^{i_N}}}{\delta_1=...=\delta^{N}=0} \label{gamma}
    % \end{align}
    \begin{equation*}
        a' = a + \sum_{1\leq|i|\leq p} \delta_1^{i_1}...\delta_N^{i_N}\gamma_{i_1,...,i_N} + O(||\delta||_\infty^{p+1}),
    \end{equation*}
    with
    \begin{equation}
        \gamma_{i_1,...,i_N} = \frac{1}{i_1!...i_N!}\evalat[\Big]{\frac{\partial^{|i|} a}{\partial\delta_1^{i_1}...\partial\delta_N^{i_N}}}{\delta_1=...=\delta_N=0} \label{gamma}.
    \end{equation}
    
    For $m\in\mathbb{N}$ different observables $A'_k$, we have
    % \begin{equation} \label{taylor_k}
    %     a'_k = a + \sum_{q=1}^p \sum_{\scriptscriptstyle i_1+...+i_N=q}\delta_{1,k}^{i_1}...\delta_{N,k}^{i_N}\gamma_{i_1,...,i_N} + O(||\delta||_\infty^{p+1})
    % \end{equation}
    \begin{equation} \label{taylor_k}
        a'_k = a + \sum_{1\leq|i|\leq p} \delta_{1,k}^{i_1}...\delta_{N,k}^{i_N}\gamma_{i_1,...,i_N} + O(||\delta||_\infty^{p+1})
    \end{equation}
    where $$||\delta||_\infty = \max_{i\in\llbracket0,N-1\rrbracket,~k\in\llbracket0,m-1\rrbracket} |\delta_{i,k}|$$
    Importantly, $\gamma_{i_1,...,i_N}$ does not depend on $k$.
    % If one wants to correct up to $p$-th order, it suffices, for each observable $A'_k$, to consider
    Now, consider, for $k\in\llbracket1,m\rrbracket$, the vector $\vec{x}_k$ consisting of monomials appearing in the expansion:
    \begin{equation} \label{x_k}
        \vec{x}_k = (\delta_{1,k}^{i_1}...\delta_{N,k}^{i_N})_{0\leq|i|\leq p}
    \end{equation}
    Let us denote $X$ the matrix of the $\vec{x}_k$'s (in columns) and $b=(1,0,...,0)^T$ (of same size as $\vec{x}_k$). When $m$ is greater than the size of $\vec{x}_k$, that is $m=O(N^p)$, one can find a solution $\vec{\lambda}=(\lambda_1,...,\lambda_m)^T$ to the linear system
    \begin{equation} \label{system}
        X\vec{\lambda}=b
    \end{equation}
    The above system is equivalent to:
    \begin{equation} \label{lambda_k}
        \sum_{k=1}^m\lambda_k\delta_{1,k}^{i_1}...\delta_{N,k}^{i_N}=0
    \end{equation}
    for all $i_1,...,i_N$ such that $1\leq|i|\leq p$, and:
    \begin{equation} \label{normalisation}
        \sum_k\lambda_k = 1
    \end{equation}
    when $i_1=...=i_N=0$.
    % The key observation is that it is always possible to find $(\lambda_1,...,\lambda_m)\in\mathbb{R}^m$ such that:
    % \begin{equation*}
    %     \sum_{k=1}^m\lambda_k\vec{x}_k=0
    % \end{equation*}
    % if we take $m$ to be larger than the size of $\vec{x}_k$, that is $m=O(N^p)$.
    % The above equation is equivalent to:
    % \begin{equation} \label{lambda_k}
    %     \sum_{k=1}^m\lambda_k\delta_{1,k}^{i_1}...\delta_{N,k}^{i_N}=0
    % \end{equation}
    % for all $i_1,...,i_N$ such that $0\leq|i|\leq p$. In particular, when $i_1=...=i_N=0$, one gets:
    % Such $\vec{\lambda}$ can be found by computing the kernel of the matrix $(\vec{x}_1,...,\vec{x}_m)$.
    % The above is equivalent to:
    % \begin{equation} \label{lambda_k}
    %     \sum_{k=1}^m\lambda_k\delta_{1,k}^{i_1}...\delta_{N,k}^{i_N}=0
    % \end{equation}
    % for all $i_1,...,i_N$ such that $1\leq|i|\leq p$.
    % By normalising, one can also impose:
    % \begin{equation} \label{normalisation}
    %     \sum_k\lambda_k = 1
    % \end{equation}
    As a result, combining Eqs. \eqref{taylor_k}, \eqref{lambda_k} and \eqref{normalisation}, one gets:
    % \begin{align*}
    %     \sum_k \lambda_k a'_k &= a + \sum_{q=1}^p \sum_{\scriptscriptstyle i_1+...+i_N=q}\left(\sum_k\lambda_k\delta_{1,k}^{i_1}...\delta_{N,k}^{i_N}\right)\gamma_{i_1,...,i_N} \\
    %       & \phantom{{}= a + \sum_{q=1}^p \sum_{\scriptscriptstyle i_1+...+i_N=q}\sum_k} + \sum_k\lambda_k O\left(||\delta||_\infty^{p+1}\right)\\
    %       &= a + \sum_k\lambda_k O\left(||\delta||_\infty^{p+1}\right)
    % \end{align*}
    \begin{align*}
        \sum_k \lambda_k a'_k &= a + \sum_{1\leq|i|\leq p}\left(\sum_k\lambda_k\delta_{1,k}^{i_1}...\delta_{N,k}^{i_N}\right)\gamma_{i_1,...,i_N} \\
          & \phantom{{}= a + \sum_{q=1}^p \sum_{\scriptscriptstyle i_1+...+i_N=q}\sum_k} + \sum_k\lambda_k O\left(||\delta||_\infty^{p+1}\right)\\
          &= a + O\left(||\lambda||_1||\delta||_\infty^{p+1}\right)
    \end{align*}
\end{proof}

This first theorem and its proof provide a deterministic and exact way to construct a $p$-th order estimate of any non-degenerate eigenvalue $a$ of the target observable (or any eigenvalue if the $A'_k$'s can be parameterised with a single variable). This estimate does not require any knowledge or evaluation of the complicated coefficients $\gamma_{i_1,...,i_p}$ of Eq. \eqref{gamma}. It also does not matter if one has restrictions on the set of $A'_k$'s they can implement (as long as it is large enough): there is indeed no condition on $(\delta_{i,k})$ to find $(\lambda_1,...,\lambda_m)$ satisfying Eq. \eqref{lambda_k}. There are however choices of $(\delta_{i,k})$ that are better than others in order to minimise the error, as will be expanded in the next subsection.
% This estimate can be constructed for any choice of implementable $A'_k$'s as there is no condition on $(\delta_{i,k})$ to find $(\lambda_1,...,\lambda_m)$ satisfying Eq. \eqref{lambda_k}: they can simply be constructed by solving a linear system of equations.

\subsection{Bounding errors} \label{section:bounding_errors}

Theorem 1 establishes the existence (and a construction in its proof) of the coefficients $(\lambda_1,...,\lambda_m)$ that are involved in the estimate of $a$. However, the error bound we derived depends on $\|\vec{\lambda}\|_1$, hence on the set of implementable observables $A'_k$: if the $A'_k$'s ($\vec{\delta}_k$'s) are almost all identical, the underlying linear system of Eq. \eqref{system} will be very poorly conditioned, and $\|\vec{\lambda}\|_1$ might explode. 
% It is therefore important to control $\vec{\lambda}$ by exploiting the fact that the choice of $\vec{\lambda}$ and $m$ is not unique.
It is therefore important to control $\vec{\lambda}$ by noting that its choice is not unique, and that some $\vec{\delta}_k$'s may lead to better solutions.

Further, controlling $\vec{\lambda}$ appears necessary to tame fluctuations of the estimated eigenvalue $a$ due to imperfect evaluations of the eigenvalues $a'_k$. 
It is particularly relevant in our problem setting as these eigenvalues will likely be measured via phase estimation, whose outcome statistically varies around the exact eigenvalue.
If each $a'_k$ is estimated with variance $\sigma^2$, that of $a$ can be expressed as
\begin{equation} \label{variance}
    \mathbb{V}\left(\sum_{k=1}^m\lambda_ka'_k\right) =  \sigma^2\sum_{k=1}^m \lambda_k^2
\end{equation}
where $\mathbb{V}$ denotes the variance.
%(depending on the number of ancillae used in the algorithm). 
Thus, having some control over $\vec{\lambda}$ would both prevent target-estimate errors and amplification of PE statistical variations.

Consequently, Theorems 2 and 3 aim at bounding $\vec{\lambda}$ in two relevant cases: when $p=1$ (first-order correction) and when $N=1$ (mono-variate case).

\begin{theorem}\label{thm:l1norm}
    Suppose that one wants to correct up to first order ($p=1$) and that $\vec{0}$ is in the convex hull of all implementable $\vec{\delta}_k$'s. It follows that there exists $\vec{\lambda}$ satisfying both Eqs. \eqref{lambda_k} and \eqref{normalisation}, and:
    \begin{equation}
        ||\vec{\lambda}||_1=1
    \end{equation}
\end{theorem}
\begin{theorem}\label{thm:l2norm}
    Suppose that the implementable observables are parameterised with a single variable ($N=1$, $\vec{\delta}_k:=\delta_k$) and that the set $\Delta$ of implementable $\delta_k$'s satisfies:
    \begin{equation}
        \exists c>0~~\forall \delta\in\Delta~~|\delta| \leq c
    \end{equation}
    \begin{equation} \label{min_dist}
        \exists d>0~~\forall (\delta,\delta')\in\Delta^2~~\delta\neq\delta'\Longrightarrow|\delta - \delta'| \geq d
    \end{equation}
    If follows that there exists $\vec{\lambda}$ satisfying both Eqs. \eqref{lambda_k} and \eqref{normalisation}, and:
    \begin{equation}
        ||\vec{\lambda}||_2 = O(1)
    \end{equation}
\end{theorem}

\begin{remark}
    Note that the first result is on the $\ell^1$-norm while the second one is on the $\ell^2$-norm. For any $\vec{\lambda}=(\lambda_1,...,\lambda_m)^T$, they are related by:
    \begin{equation} \label{eq:norms}
        ||\vec{\lambda}||_2 \leq ||\vec{\lambda}||_1 \leq \sqrt{m}||\vec{\lambda}||_2
    \end{equation}
\end{remark}

The proof of these theorems can be found in the appendix. Theorem \ref{thm:l1norm} gives the configuration of the $\vec{\delta}_k$'s that provides a $\vec{\lambda}$ achieving minimal $\|\vec{\lambda}\|_1$. Indeed, given the normalisation condition $\sum\lambda_k=1$:
\begin{equation*}
    ||\vec{\lambda}||_1 \geq \sum_k\lambda_k = 1
\end{equation*}
and
\begin{equation*}
    ||\vec{\lambda}||_1 = 1~~\Longleftrightarrow~~\forall k~\lambda_k\geq0
\end{equation*}
This exactly means that $\vec{0}$  is in the convex hull of all implementable $\vec{\delta}_k$'s. Also note that this configuration brings the $\ell^2$-norm closer to its minimum $1/\sqrt{m}$, which is achieved when all $\lambda_k$'s are equal and positive (using Cauchy-Schwarz inequality).

As for Theorem \ref{thm:l2norm}, it shows, in a simpler case than the general one, that the chosen $\vec{\delta}_k$'s must be separated enough to build the eigenvalue estimator with small variance. Otherwise, the linear system of Eq. \eqref{lambda_k} will have a high condition number, thus yielding high $\lambda_k$’s.

It seems harder to assert anything for all the other cases than $p=1$ or $N=1$. In particular, the latter cannot be adapted to $N\geq2$ as a much more complex matrix than the Vandermonde matrix would appear in the proof at Eq. \eqref{vandermonde}. It would not be clear where the singular values of that matrix vanish, compared to the well-known case of the Vandermonde matrix: it is thus harder to give an explicit condition such as the one of Eq. \eqref{min_dist}. In practice however, the numerical simulations of the next section show that the eigenvalue estimate remains robust even when $p\geq2$ and $N\geq2$.

% However, one can note that in the most general case, these singular values only cancel on isolated points (as the roots of a polynomial). These points are undetermined, yet when picking a set of $\vec{\delta_k}$'s for the construction of the eigenvalue estimate, there is a strictly 0 probability of picking 

\section{Numerical results}

% In the following, we will note $V=H-H'$ the difference of $A$ and $A'$, and write its decomposition into the relevant operator basis as (note the superscript to facilitate later notations):
% $$V=\sum_{i=0}^{N-1}\delta_iO_i$$
% $N$ will hence refer to the number of non-zero terms in $V$.

% \subsubsection{Method 1}

% The first more straightforward method consists of expanding the eigenvalues $E$ of $A$ as a function of the eigenvalues $E'$ of $A'$:

% \begin{align}
%     E &= E' + \bra{\psi'}V\ket{\psi'} + O(V^2)\\
%       &= E' + \sum_i \delta_i\bra{\psi'}O_i\ket{\psi'} + O(\delta^2)
% \end{align}
% with $\ket{\psi'}$ the ground-state of $A'$, generated as an output of phase estimation on $\mathcal{W}(H')$.

% Since the $\delta_i$'s are known and $E'$ is obtained from phase estimation on $\mathcal{W}(H')$, the mitigation strategy then simply consists of estimating the quantities $\bra{\psi'}O_i\ket{\psi'}$. This can be done by performing phase estimation a large number of times, so as to generate enough copies of $\ket{\psi'}$ to estimate the aforementioned average values.

In this section, we test our error mitigation protocol on one-dimensional Ising and Heisenberg XYZ  Hamiltonians with $n$ spins:
\begin{align}
    H_{\text{Ising}} &= \sum_{i=1}^n a_iZ_i + \sum_{i=1}^n b_iX_iX_{i+1}\\
    % H_{\text{Heisenberg}} &= \sum_{i=1}^n a_iZ_i + \sum_{i=1}^n (b_iX_iX_{i+1} + c_iY_iY_{i+1} + d_iZ_iZ_{i+1} \\
    H_{\text{XYZ}} &= \sum_{i=1}^n a_iZ_i + \sum_{i=1}^n (b_iX_iX_{i+1} + c_iY_iY_{i+1} \label{Heisenberg} \\
    & \phantom{{}=\sum_{i=1}^n a_iZ_i + \sum_{i=1}^n (b_iX_iX_{i+1}} + d_iZ_iZ_{i+1}) \notag
\end{align}
where an index $n+1$ is taken mod $n$ hence refers to spin 1.
The coefficients $a_i$, $b_i$, $c_i$ and $d_i$ are chosen randomly from a uniform distribution between 0 and 1, but only once for each chain length.
Our aim is to estimate the ground-state energy of the target Hamiltonian $H$, via the measurement of the ground-state energy of implementable Hamiltonians $H'_k$. The set of implementable Hamiltonians varies from subsection to subsection, depending on the assumptions on the implementation of $\mathcal{W}(H)$. Our numerical simulations do not use any quantum circuit simulator. The implementable Hamiltonians are directly diagonalised, which is achievable for small enough spin chains. In the first subsection, phase estimation errors are ignored.

\subsection{$p$-th order mitigation of Trotter errors}

First, we verify the efficiency of our protocol in the simplest case: when the set of implementable Hamiltonians is parameterised by a single parameter ($N=1$). Such a case is for instance relevant for trotterised Hamiltonians, where this parameter is the timestep $\delta t$ (Eq. \eqref{trotter}). We use the XYZ model in this demonstration. In this case, the Hamiltonian can be split into two sub-Hamiltonians $H_A$ and $H_B$ obtained by respectively summing for even and odd $i$ in Eq. \eqref{Heisenberg} \cite{yi2021spectral}. By compiling $\mathcal{W}(H')=\left(\mathrm{e}^{-\mathrm{i}H_A\delta t } \mathrm{e}^{-\mathrm{i}H_B\delta t}\right)^{t/\delta t} = \mathrm{e}^{-\mathrm{i}H't}$, the effectively implemented Hamiltonian $H'=\tilde{H}(\delta t)$ thus reads:
\begin{equation*}
    H'=\tilde{H}(\delta t)=\frac{\mathrm{i}}{\delta t}\mathrm{log}\left(\mathrm{e}^{-\mathrm{i}H_A\delta t } \mathrm{e}^{-\mathrm{i}H_B\delta t }\right)
\end{equation*}

Let us denote $N_\text{Trotter,max}$ the maximum number of Trotter steps a device can tolerate. While the best possible $H'$ is obtained by using $N_\text{Trotter,max}$ Trotter steps, multiple $H'_k$ can be implemented by reducing the number of Trotter steps, effectively increasing the algorithmic errors \cite{Endo_2019}. Another possibility is to note that $\delta t$ can be chosen to be negative:
% If the ideal Hamiltonian is obtained in the limit $\delta t \rightarrow 0$, the implementable Hamiltonians $H'$ can only tolerate a minimum timestep $|\delta t_{\mathrm{min}}|=t/N_\text{Trotter,max}$, with $t$ a free parameter (total time) and $N_\text{Trotter,max}$ the maximum number of Trotter steps. Multiple $H'_k$ can be implemented by using less Trotter steps than $N_\text{Trotter,max}$, effectively increasing the algorithmic errors \cite{Endo_2019}, but also by noticing $\delta t$ can be chosen to be negative:
\begin{align*}
    \tilde{H}(-\delta t) &= -\frac{\mathrm{i}}{\delta t}\mathrm{log}\left(\mathrm{e}^{+\mathrm{i}H_A\delta t } \mathrm{e}^{+\mathrm{i}H_B\delta t }\right) \\
    &= \frac{\mathrm{i}}{\delta t}\mathrm{log}\left(\left(\mathrm{e}^{+\mathrm{i}H_A\delta t } \mathrm{e}^{+\mathrm{i}H_B\delta t }\right)^{-1}\right) \\
    &= \frac{\mathrm{i}}{\delta t}\mathrm{log}\left(\mathrm{e}^{-\mathrm{i}H_B\delta t } \mathrm{e}^{-\mathrm{i}H_A\delta t }\right)
\end{align*}
Hence, if $\tilde{H}(\delta t)$ is implementable, $\tilde{H}(-\delta t)$ surely is too, as it just involves swapping $H_A$ and $H_B$. This trick allows one to halve the level of increase of algorithmic error that is needed to implement enough $H'_k$’s. For a given level of error mitigation $p$, Eq. \eqref{system} is a system of $p+1$ linear equations, hence has a non-zero solution $(\lambda_1,...,\lambda_m)$ for $m\geq p+1$. We thus choose the set $\Delta$ of implemented $\delta t_k$ to be:
\begin{equation}
    \Delta = \{k\times\delta t_{\mathrm{min}},~k \in\llbracket-p/2,p/2+1\rrbracket\setminus\{0\}\}
\end{equation}
After obtaining $\vec{\lambda}$ corresponding to these $\delta t_k$'s, we estimate the ground state energy of the original Hamiltonian via Eq.~\eqref{master}, using the ground state energies of the $H_k$'s computed by exact diagonalisation.

\begin{figure}
    \centering
    \includegraphics[width=\linewidth]{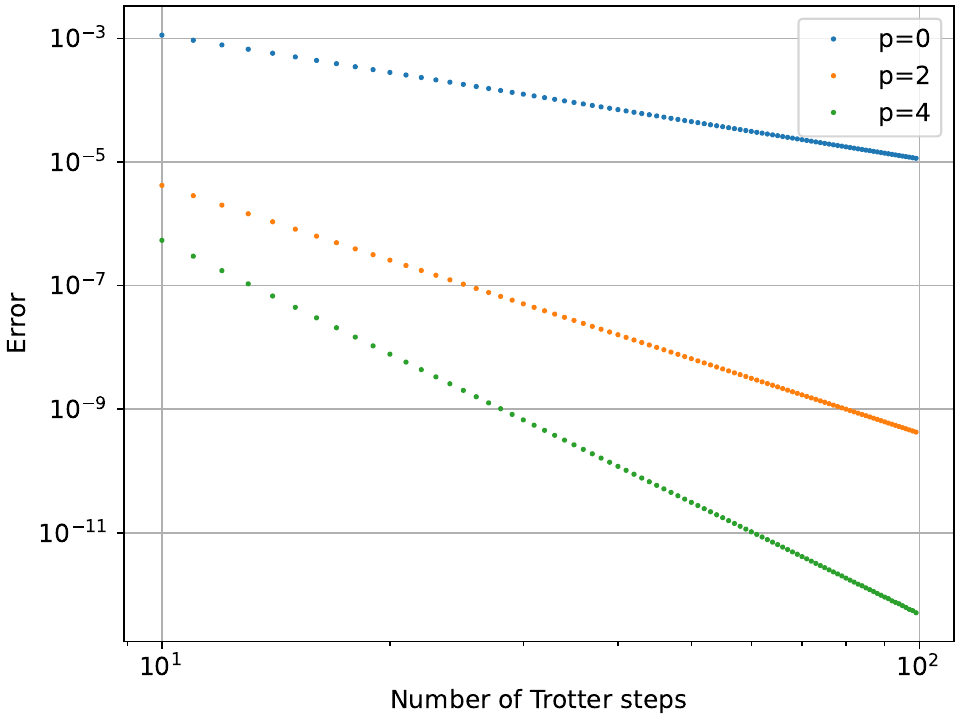}
    \caption{Error in the ground-state energy estimation of an XYZ Hamiltonian with $n=6$ particles implemented by Trotterisation, plotted against the maximum number of Trotter steps. The coefficients of the Hamiltonian are randomly chosen before all the experiments. Each colour corresponds to a different order $p$ of error mitigation, $p=0$ meaning no error mitigation is performed.}
    \label{fig:trotter}
\end{figure}

The results of such an approach are presented in Fig.~\ref{fig:trotter}, where the error in the ground-state energy estimation is plotted against $N_\text{Trotter,max}$ for $p=0$ (no error mitigation), $p=2$ and $p=4$. The straight lines in log-log scale indicate a power law:
\begin{equation*}
    \text{error} = \left(\frac{1}{N_\text{Trotter,max}}\right)^\alpha
\end{equation*}
For a given $p$, one would expect a $p$-th order estimate, hence $\alpha=p+1$. However, a linear regression for each plot of Fig. \ref{fig:trotter}, for $p=0$, 2 and 4, respectively suggests $\alpha=2$, 4 and 6. This indicates that the ground state has no odd order component in its Taylor expansion: correcting for, say, second order, one only retains a fourth order error. This feature was noted in \cite{yi2021spectral} for the first-order component, given by $E^{(1)} = \bra{0}H'-H\ket{0}$. They noticed that after Trotterisation, $H'-H$ was off-diagonal for a wide category of Hamiltonians, including any Hamiltonian with nearest-neighbour interactions like $H_{\mathrm{XYZ}}$ employed here. The results of Fig. \ref{fig:trotter} hence suggest that this is not only true for first order, but for all odd orders. For even $p$, it thus becomes possible to obtain a $p+1$-th order estimate of the ground state with only $p+1$ approximate ground-state energy estimations.

Further, a natural question to ask is how robust this protocol is to random errors. Indeed, if our scheme does guarantee the suppression of algorithmic errors, one would not want random errors to accumulate in the process. To model these random errors, we will assume in the following that the QPE circuit is affected
% by depolarising noise at the logical level. More precisely, each , controlled-U or SWAP 
, at the logical level, by circuit-level depolarising noise of strength $q$ (in the sense explained in \cite{faizan2023simulation}, thus excluding noise within the circuit preparing $U$, which we will comment on later). As shown in \cite{faizan2023simulation}, this impacts the output $E$ of QPE in two ways: its average value $\bar{E}$ exponentially approaches $0.5$ when $q$ increases; and its standard deviation $\Delta E$ increases and quickly saturates. Because we are still interested in a fault-tolerant setup, the strength $q$ of the depolarising noise can be assumed to be small, say under $10^{-5}$ for early fault-tolerance. 

To evaluate the impact of the noise, we decide to plot the average Euclidean distance between our estimate $\tilde{E}$ for the ground-state energy of $H$ and its exact value $E$: $\sqrt{\mathbb{E}((\tilde{E}-E)^2)}$, where $\mathbb{E}$ denotes the expectation value. This quantity is evaluated by rerunning the same experiment as in Fig. \ref{fig:trotter} but with additional random noise: the ground state energies of the $H'_k$'s are now sampled according to normal distributions extracted from Table 1 and Fig. 2 of \cite{faizan2023simulation}. Because of the randomness of the current experiment, each data point is also averaged over $N_{\text{runs}}=1000$.

For a low number of Trotter steps, \textit{i.e.} when the level of the algorithmic errors is well above the level of the random noise, the data from Fig. \ref{fig:trotter} is unchanged. However, because of the presence of the random noise that our protocol cannot correct, the error rate does not exponentially decrease towards 0 but now saturates. The floor value can be computed exactly:
\begin{align}
    \sqrt{\mathbb{E}((\tilde{E}-E)^2)} &= \sqrt{\mathbb{V}(\tilde{E}-E) + \mathbb{E}(\tilde{E}-E)^2} \nonumber\\
        &= \sqrt{||\lambda||_2^2 \Delta E^2 + (\bar{E}-E)^2} \label{eq:floor_random_noise}
\end{align}
where we used Eq. \ref{variance} between the first and second line, and assumed $\mathbb{E}(\tilde{E})=\bar{E}$, where $\bar{E}$ denotes the erroneous average value from the noisy QPE, without any additional algorithmic error in case of the floor value. We observe that as the order of error mitigation $p$ increases, the floor value increases too, as $||\lambda||_2$ does. This increase is not dramatic though, as $||\lambda||_2=1$, 1.4 and 2.4 for respectively $p=0$, 2 and 4. Besides, before this floor is reached, the error suppression promised by our error mitigation technique is still achieved: for instance, in all noise cases and on an early-fault tolerant device where only 10 Trotter steps may be reliably implementable, our error mitigation protocol always outperforms the no-mitigation case, still at the very low cost of 3 or 5 ground state energy evaluations.

\begin{figure}
    \centering
    \includegraphics[width=\linewidth]{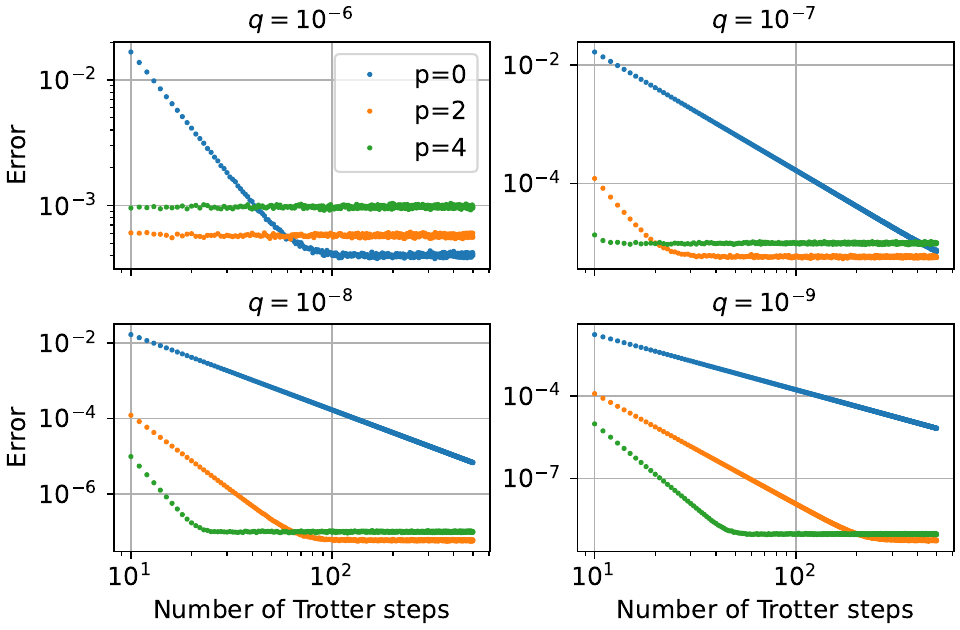}
    \caption{Average Euclidean distance between the estimate $\tilde{E}$ for the ground-state energy of $H$ and its exact value $E$, $\sqrt{\mathbb{E}((\tilde{E}-E)^2)}$, in the presence of circuit-level depolarising noise on the logical qubits. The same set of parameters as in the experiment of Fig. \ref{fig:trotter} is used. Each data point corresponds to an average value over $N_{\text{runs}}=1000$ experiments. Each subplot corresponds to a different level $q$ of depolarising noise.}
    \label{fig:trotter_with_depolarising}
\end{figure}

Finally, note that we here derived a very simple noise model from \cite{faizan2023simulation}.
In particular, we assumed a normal distribution of the random errors (which was not proven in \cite{faizan2023simulation}). This should however have no impact on the data of Fig. \ref{fig:trotter_with_depolarising} as after averaging over a sufficiently high $N_{\text{runs}}$, the output should only depend on $\bar{E}$ and $\Delta E$ and not on the other characteristics of the noise (see Eq. \ref{eq:floor_random_noise}). Besides, we did not include any depolarising channel within the circuit preparing $U$. As a result, increasing the number of Trotter steps here has no influence on the total random noise, which should normally increase as the depth of the quantum circuit grows as well.
% In particular, we did not assume that increasing the number of Trotter steps would increase the total random noise, which should be the case as the depth of the quantum circuit, hence the number of depolarising channels, would grow as well.
But this refined model would only make the floor non-constant without changing the main observations of the previous plot: the error suppression saturates when random noise is added, but the floor value barely changes when our error mitigation scheme is used (at most multiplied by 2.4 for 4th order mitigation). The effect of the random noise could even be lowered if $||\lambda||_2$ could be chosen to be smaller than 1: this is difficult in the case of singly-parameterised Hamiltonians as the parameter space is small, but will become easier in the next sections focusing on multi-parameter Hamiltonians. Besides, above the floor, the random noise does not affect the efficiency of our protocol and algorithmic errors are indeed suppressed. In the rest of the paper, we will thus assume that we are in the regime where random errors are well-below algorithmic errors, thereby neglecting them. $||\lambda||_2$ will also always be engineered to be lower than 1, guaranteeing no random error amplification.

\subsection{$p$-th order mitigation of qubitised Hamiltonians}

The previous subsection addressed the case of singly-parameterised observables, with the example of trotterised Hamiltonians. In this subsection, we tackle the case of multiple parameters ($N>1$), with the example of qubitised Hamiltonians. As explained in Eq. \eqref{qubitisation}, given a target Hamiltonian expressed as a linear combination of unitaries $H=\sum c_iP_i$, the effectively implemented Hamiltonians by qubitisation are $H'_k=\sum c'_{i,k}P_i$. We here absorb the renormalisation constants and assume $\sum_i c_i = \sum_i c'_{i,k} = 1$.

For all $i$ and $k$, the coefficients $c_i$ and $c'_{i,k}$ are between 0 and 1 and close to each other. Following the procedure designed in \cite{Babbush_2018}, $c'_{i,k}$ is chosen to be a $\mu$-bit number close to $c_i$ (either the closest, second closest or third closest, in order to have multiple implementable $H'_k$), before normalising all $c'_{i,k}$'s:
\begin{align*}
    c'_{i,k} &= \text{round}(2^\mu c_i)/2^\mu + \epsilon_{i,k},~~\epsilon_{i,k} \in \{-1/2^\mu,0,1/2^\mu\}\\
    c'_{i,k} &\leftarrow \frac{c'_{i,k}}{\sum_i c'_{i,k}}
\end{align*}
where $\text{round}(x)$ returns the closest integer to the float $x$. $(\epsilon_{i,k})$ should be chosen such as to minimise the norm of $\vec{\lambda}=(\lambda_1,...,\lambda_m)^T$. In the present study, we did not try to optimise this choice: for a given set of $\epsilon_{i,k}$'s, one can compute a $\vec{\lambda}$ of minimum $\ell^2$-norm satisfying both Eqs. \eqref{lambda_k} and \eqref{normalisation}.
Our approach is thus to repeatedly choose at random the set of $\epsilon_{i,k}$'s till this $\vec{\lambda}$ satisfies a certain condition (norm less than 1 or positive coefficients).
% The set of $\epsilon_{i,k}$'s can thus be repeatedly chosen at random till this $\vec{\lambda}$ satisfies a certain condition (norm less than 1 or positive coefficients).
If the condition cannot be met, the number $m$ of implemented Hamiltonians is incremented, thereby giving more freedom in the choice of $\vec{\lambda}$.

% As we stated in Section \ref{section:bounding_errors}, the error is proportional to the

In addition, we here consider statistical errors arising from the use of the phase estimation algorithm. The ground-state energy of an implementable Hamiltonian is obtained by applying PE to $\mathcal{W}(H)=\mathrm{e}^{-\mathrm{i}\mathrm{arccos}(H)}$. If the target eigenvalue is $\mathrm{e}^{-\mathrm{i}2\pi\phi}$, the output distribution is given by:
\begin{equation} \label{eq:pe}
    \mathbb{P}(y=y_0,...,y_{q-1})=\frac{1}{2^{2q}}\left|\frac{1-\mathrm{e}^{2\mathrm{i}\pi\delta(y)2^q}}{1-\mathrm{e}^{2\mathrm{i}\pi\delta(y)}}\right|^2
\end{equation}
where $q$ is the number of ancillae used in phase estimation, $\mathbb{P}(y=y_0,...,y_{q-1})$ is the probability of observing the output state $\ket{y_0,...,y_{q-1}}$, and $\delta(y)=\phi-y=\phi-\sum_{i=0}^{q-1} y_i/2^{i+1}$. Finally, the energy is:
\begin{equation} \label{eq:cos}
    E=\mathrm{cos}(2\pi y)
\end{equation}
We do not take into account additional fluctuations stemming from the sampling of other eigenvalues than the ground state (arising when the input state of PE is not exactly the ground state of the implemented Hamiltonian). This is a fair assumption if the ground and first excited states are separated enough: in this case, any outlier in the sampled data can be identified and ignored, and PE can be performed again.

\begin{figure}
    \centering
    \includegraphics[width=\linewidth]{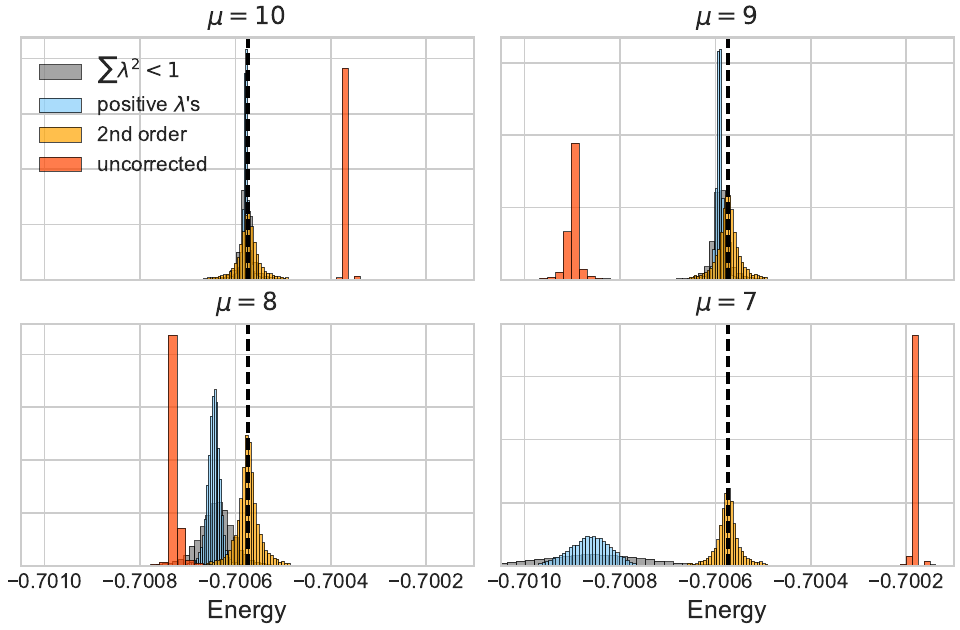}
    \caption{Distribution of the estimated ground-state energy of an Ising Hamiltonian with 8 particles and random coefficients, with phase estimation errors taken into account. PE is applied to $\mathcal{W}(H)=\mathrm{e}^{-\mathrm{i}\mathrm{arccos}(H)}$, implemented by qubitisation. This requires the preparation of a state $\ket{G}$, which can only be done up to a certain precision proportional to $1/2^\mu$. The red data corresponds to the raw output obtained from PE, while the others are post-processed with our error mitigation scheme: gray and blue correspond to first order correction with two different conditions on $\vec{\lambda}$, and yellow corresponds to second order.}
    \label{fig:binary}
\end{figure}

The procedure of the numerical experiment performed here is thus as follows.
First, we repeatedly generate $(\epsilon_{i,k})$ randomly to construct sets $\{H_k'\}$ of implementable Hamiltonians, and stop when sufficiently good $\vec{\lambda}$ is obtained.
Second, we compute the ground state energies of the retained $H_k'$'s by exact diagonalisation.
Third, we sample once the output of PE from \eqref{eq:pe} and compute the associated energies with \eqref{eq:cos}, using $q=16$ ancillae.
Fourth, we calculate the target eigenvalue estimator using $\vec{\lambda}$ and the sampled energies.
Given the inherent randomness of this protocol, the four steps above are then repeated 10,000 times, hence giving 10,000 different estimates of the target energy, which are represented with histograms in Fig.~\ref{fig:binary}.

This figure is thus obtained using such a protocol for the ground-state energy estimation of an Ising chain with $n=8$ particles. This means 16 unitaries in the decomposition of $H$, which translates into $N=16$ parameters $\delta_i = c'_i-c_i$. Each subplot corresponds to a different level of noise (high $\mu$ meaning low noise) and the different colours to different error mitigation strategies: gray is first order and enforcing $||\vec{\lambda}||_2<1$, blue is first order and enforcing $\lambda_k>0$ for all $k$ (which leads to $\|\vec{\lambda}\|_1=1$), and yellow is second order and enforcing $||\vec{\lambda}||_2<1$.
The target ground-state energy is indicated with a black vertical dashed line.
The ground-state energy one would get by applying phase estimation to the best implementable Hamiltonian $H'_0$ ($\epsilon_{i,0}=0$ for all $i$) is plotted in red: it follows the distribution given by Eqs. \eqref{eq:pe} and \eqref{eq:cos}. Because of the algorithmic errors, there is a bias between the target value (vertical dashed line) and the data.
This bias is corrected by the application of our error mitigation strategies. As expected, first order correction (blue and gray) is sufficient to cancel the bias at low enough algorithmic errors ($\mu=10$) while second order correction (yellow) offers greater correction at any plotted $\mu$.
Besides, one can observe that the variance of the corrected data increases with the strength of the errors (decreasing $\mu$) as the fluctuations in $\epsilon_{i,k}$ increase. Note that even for the same $p$, the variance of the corrected data depends on the error mitigation strategy, as it is also proportional to $||\vec{\lambda}||_2$. As expected from Theorem 2, choosing the $\epsilon_{i,k}$'s such that all $\lambda_k$'s are positive is most optimal: this translates into a minimum $\ell^1$-norm and brings the $\ell^2$-norm closer to its minimum. This results in a narrower distribution for the blue than for the gray data.

% This is particularly visible in the bottom right plot where the gray histogram is broader than the blue one: in the gray data $||\vec{\lambda}^*||<1$ is enforced, while in the blue one all coefficients of $\vec{\lambda}^*$ are positive, bringing its $\ell^2$-norm closer to its lower bound $1/\sqrt{m}$ (Eq. \eqref{eq:norms} with $||\vec{\lambda}^*||_1=1$).

It hence naturally appears that increasing the complexity of the error mitigation method (gray to blue to yellow) improves the quality of the output. This obviously comes at a cost which is studied in the next subsection.

\subsection{Cost of the proposed scheme}

% The last subsection established the number of $m$ of approximates one would need to obtain a $p$-th order estimate in the case of singly parameterised Hamiltonians ($N=1$). In this subsection we will more generally estimate how the error evolves with the number of times phase estimation is performed.

The cost of our protocol can be estimated by the number of times $n_{\mathrm{PE}}$ that phase estimation must be performed. In order to obtain a $p$-th order estimate of a given eigenvalue $a$, one must be able to find coefficients $(\lambda_1,...,\lambda_m)$ satisfying Eqs. \eqref{lambda_k} and \eqref{normalisation}. This is always possible if $m$ is greater than the size $s$ of the vector $\vec{x}_k = (\delta_{1,k}^{i_1}...\delta_{N,k}^{i_N})_{0\leq|i|\leq p}$, which can be calculated as:
\begin{equation}
    s = \sum_{q=0}^p {N+q-1\choose N-1}
\end{equation}
where we used that the number of tuples $(i_1,...,i_N)\in\mathbb{N}^N$ of fixed sum $q$ is ${N+q-1\choose N-1}$. In some cases though, the $\vec{x}_k$'s may have some internal linear dependencies, thereby reducing the rank of their matrix: less $\lambda_k$'s would thus be required. This is for example the case for normalised qubitised Hamiltonians, for which $\sum_i c_i = \sum_i c'_{i,k} = 1$, meaning that, for all $k$, $\sum_i \delta_{i,k} = 0$. This identity reduces the rank by 1 for $p=1$, and creates even more linear dependencies for higher $p$.

\begin{figure}
    \centering
    \includegraphics[width=\linewidth]{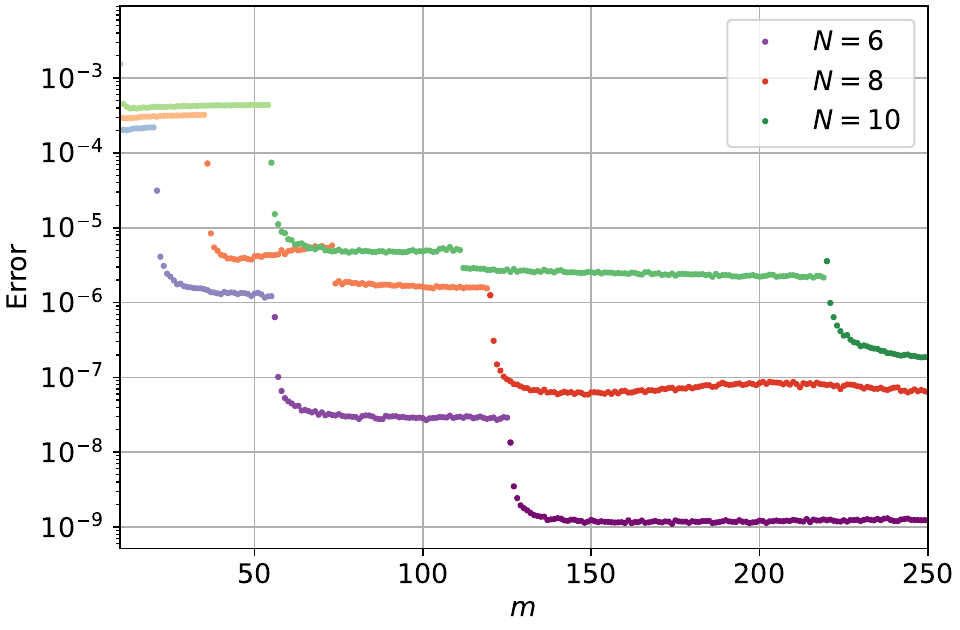}
    \caption{Error in the ground-state energy estimation of a qubitised Ising Hamiltonian with $N=6$, 8 and 10 terms, plotted against the number $m$ of distinct Hamiltonians implemented. Each distinct colour corresponds to a different number of terms in the Hamiltonian, while the slight changes of shade happening when the error drops coincide with a change in error mitigation order $p$.}
    \label{fig:vary_m}
\end{figure}

Besides, it is important to note that, for a given $p$, choosing the lowest possible $m$ allowing one to construct $(\lambda_1,...,\lambda_m)$, which we denote by $m_{p,\mathrm{min}}$, is not the most clever choice.
Indeed, as the parameter $m$ (\textit{i.e.} the number of variables of the linear system $X\vec{\lambda}=b$ of Eq. \eqref{system}) increases, the solution space also expands, potentially permitting a smaller minimum norm for $\vec{\lambda}$. Consequently, this leads to a reduced error since the bound on $\delta a$, as stated in Theorem 1, depends on $||\vec{\lambda}||_1$.
% This is because there are more solutions to the linear system $\sum\lambda_k\vec{x_k}=0$, hence potentially allowing a lower minimum norm for $\vec{\lambda}$, hence a lower error (as the bound on $\delta a$ depends on $||\vec{\lambda}||_1$ in Theorem 1).

We thus conduct a final experiment to gain insights on the optimal choice of $m$.
In Fig. \ref{fig:vary_m}, we plotted the error in the ground-state energy estimation of a qubitised Ising Hamiltonian of varying length, using the same protocol as in the previous subsection but with only one random choice of $\epsilon_{i,k}$ (without trying to meet any additional condition on $\vec{\lambda}$). Each data point corresponds to the average distance between the estimator and the target eigenvalue, over 10,000 experiments. The highest possible order of error mitigation $p$ is always chosen, \textit{i.e.} whenever $m$ exceeds the rank of the matrix $X$ of the $\vec{x}_k$'s. An increase in $p$ is, as expected, characterised by a drop in the error, and is highlighted by a change of shade of the same colour.
This drop is however relatively smooth: Fig. \ref{fig:vary_m} shows that choosing $m=m_{p,\mathrm{min}}$ fails to significantly improve the results of the $(p-1)$-th order mitigation. In turn, a slightly higher $m$ yields results that faithfully achieve $p$-th order mitigation. This increase of $m$ can however remain very moderate, only a few units (say, 10), as for a given $p$ the error quickly stabilises. Therefore, to achieve minimum error for a given $p$, one can set:
% Indeed, Fig. \ref{fig:vary_m} shows that, while at the minimal $m$ for a given $p$ fails to improve from the results of $(p-1)$th order mitigation, the drop in the error is relatively smooth and choosing a slightly higher $m$ than the minimal yields results that faithfully achieve $p$th order mitigation. This increase of $m$ can however remain very moderate, only a few units (say, 10), as for a given $p$ the error quickly stabilises. Therefore, to achieve minimum error for a given $p$, one can set:
\begin{equation}
    m = m_{p,\mathrm{min}}+10 = O(s)
\end{equation}

Assuming that the initial state used for phase estimation has an overlap $\beta$ with the real ground state and that the first excited state is separated enough from the ground state to differentiate them, we conclude that the number of times $n_{\mathrm{PE}}$ phase estimation must be performed can be written as:
\begin{equation*}
    n_{\mathrm{PE}} = O(m/\beta) = O(s/\beta)
\end{equation*}
Thus:
\begin{equation}
    n_{\mathrm{PE}} = 
    \begin{cases}
        O(N^p/\beta),\text{ if }N\neq1\\
        O(p/\beta),\text{ if }N=1
    \end{cases}
\end{equation}
The number of times phase estimation must be performed is thus polynomial in the number of parameters the Hamiltonian depends on. In the case of singly-parameterised Hamiltonians, this number is even linear in the desired order of mitigation, enabling powerful error reduction at very low cost.

\section{Discussion}

In this paper, we thus presented an extension of Richardson extrapolation, introducing several key novelties. Specifically, these include the mitigation of eigenvalues rather expectation values, and the use of a multi-parameter formalism without which Richardson extrapolation would fail to work for qubitised Hamiltonians.

% In this paper, we thus presented a new error mitigation protocol for the problem of estimating eigenvalues of given observables $A$. Our method is an extension of Richardson extrapolation, introducing key novelties

Our scheme targets algorithmic errors, that is errors arising from approximations in the quantum algorithm implementing the unitary passed to phase estimation (\textit{e.g.} $\mathcal{W}(H)=\mathrm{e}^{-\mathrm{i}Ht}$). After proving the theoretical performance of the proposed method, we numerically confirmed its efficiency and low resource requirements. In particular, we showed that for some relevant cases, such as that of trotterised Hamiltonians, a $p$-th order estimate of the ground-state energy can be obtained with order of $p$ (independent) uses of phase estimation, thereby drastically reducing the error at very low cost. On top of this, we conducted a novel theoretical and numerical study aiming at better understanding how the chosen noise parameters and number $m$ of implemented observables impact the final error rate (via $||\vec{\lambda}||_2$). As such, we examined how these parameters can be fine-tuned to minimise the error. In particular, we extended the study of zero-noise extrapolation to a more general framework where negative parameters are allowed, showing that they can indeed be conducive to more optimal results.

% As for multi-parameter observables such as qubitised Hamiltonians, we conducted novel theoretical and numerical studies aiming at 

Although our protocol can be applied to the estimation of the eigenvalues of any observable, it is however important to note that it is only effective against fully known errors. Indeed, our scheme entirely relies on the full knowledge of the parameters $(\delta_{i,k})$ the implementable Hamiltonians depend on. As it might be unrealistic to assume that random errors will entirely be suppressed in early-stage fault-tolerant quantum computers, we however showed that additional random noise, albeit non correctable by our protocol, will not accumulate. This justifies the robustness and usability of our scheme even in the very first fault-tolerant devices.

As for future directions, our method could be improved by giving a more systematic way to choose \textit{good} $\delta_{i,k}$'s out of the set of implementable ones, so as to minimise the norm of $\vec{\lambda}=(\lambda,...,\lambda_m)^T$. In the previous subsections, the $\delta_{i,k}$'s were chosen randomly till some conditions on $\vec{\lambda}$ were satisfied. If this yielded good results (in particular still guaranteed correction at the desired order $p$), a more optimised approach could help minimise the prefactor $||\vec{\lambda}||_1$ in the error term. As a result, one would understand more systematically the number $m$ of $\lambda_k$'s that are needed to obtain, for a given $p$, a minimal error.

\section*{Acknowledgements}
This work was supported by MEXT Quantum Leap Flagship Program (MEXT Q-LEAP) Grant No. JPMXS0120319794 and JST COI-NEXT Grant Number JPMJPF2014. 
KM is supported by JST PRESTO Grant No. JPMJPR2019 and JSPS KAKENHI Grant No. 23H03819.
\bibliography{ref}

\appendix

\newpage

\section{Proof of Theorem 2}

\begin{proof}
        The case ($p=1$) is quite straightforward. First note that for any $\vec{\lambda}$ verifying Eq. \eqref{normalisation}, $||\vec{\lambda}||_1 \geq \sum_k\lambda_k\geq1$. Besides, if 0 is in the convex hull of all implementable $\vec{\delta}_k$'s, then:
    \begin{equation}
        \exists (\lambda_1,...,\lambda_m)\in\mathbb{R}_+^m~~\sum_k \lambda_k\delta_k = 0
    \end{equation}
    with: $$\sum_k\lambda_k = 1$$
    These two equations are what Eqs. \eqref{lambda_k} and \eqref{normalisation} reduce to in the case $p=1$.  This choice of $\vec{\lambda}$ achieves the lower bound:
    \begin{equation}
        ||\vec{\lambda}||_1 = \sum_k|\lambda_k| = \sum_k\lambda_k = 1
    \end{equation}
\end{proof}

\section{Proof of Theorem 3}

\begin{proof}
    This more works. For $N=1$, Eq. \eqref{lambda_k} becomes:
    \begin{equation*}
        \forall i\in\llbracket1,p\rrbracket~~\sum_{k=1}^m\lambda_k\delta_k^i = 0
    \end{equation*}
    Thus, one can note that both Eqs. \eqref{lambda_k} and \eqref{normalisation} can be combined using the rectangular Vandermonde matrix $V=(\delta_k^i)_{i,k}$ and the vector $b=(1,0,...,0)^T$ with $m-1$ zeros:
    \begin{equation} \label{vandermonde}
        V\vec{\lambda} = b
    \end{equation}
    The first equation of this linear system is Eq. \eqref{normalisation} and the others are Eq. \eqref{lambda_k}. A solution $\vec{\lambda}^*$ of minimum $\ell^2$-norm of this system satisfies:
    \begin{equation}
        ||\vec{\lambda}^*||_2 = ||V^+b||_2
    \end{equation}
    with $V^+$ the Moore–Penrose inverse of $V$. In particular:
    \begin{equation} \label{eq:bound_norm}
        ||\vec{\lambda}^*||_2 \leq ||V^+||_2||b||_2 = ||V^+||_2
    \end{equation}
    Using the singular value decomposition $V=U\Sigma V^\dagger$, we obtain:
    \begin{equation}
        ||V^+||_2 = ||\Sigma^+||_2 = \max \{1/\mu, \mu\in\mathcal{S}\setminus\{0\}\}
    \end{equation}
    where $\mathcal{S}$ denotes the singular values of $V$. Besides, it is well-known that the square Vandermonde matrix $\tilde{V}(\delta_1,...,\delta_p)$ is invertible if and only if for all $i\neq j$, $\delta_i\neq\delta_j$. As a result, the rectangular Vandermonde matrix $V$ is full-rank, hence has non-zero singular values only, if and only if for all $i\neq j$, $\delta_i\neq\delta_j$. By continuity of the singular values over $(\delta_1,...,\delta_m)$ on the compact space $\Delta$, it follows that $||V^+||_2$ is bounded, hence $||\vec{\lambda}^*||_2$ too.
\end{proof}

\end{document}